\documentstyle[12pt]{article}
\input epsf.tex
\newcommand{\beq}{\begin{equation}}
\newcommand{\eeq}{\end{equation}}
\newcommand{\beqa}{\begin{eqnarray}}
\newcommand{\eeqa}{\end{eqnarray}}
\newcommand{\beqar}{\begin{eqnarray*}}
\newcommand{\eeqar}{\end{eqnarray*}}
\renewcommand{\a}{\alpha}
\renewcommand{\b}{\beta}

\newcommand{\e}{\epsilon}
\newcommand{\g}{\gamma}
\newcommand{\G}{\Gamma}

\newcommand{\ka}{\kappa}
\renewcommand{\l}{\lambda}

\newcommand{\na}{\nabla}
\renewcommand{\O}{\Omega}

\renewcommand{\t}{\theta}
\newcommand{\z}{\zeta}
\newcommand{\cA}{{\cal A}}
\newcommand{\cl}{\ell}

\newcommand{\eg}{{\it e.g.,}\ }
\newcommand{\ie}{{\it i.e.,}\ }
\newcommand{\hphi}{\hat{\phi}}

\newcommand{\hh}{\hat{h}}
\newcommand{\hs}{\hat{s}}
\newcommand{\hB}{\hat{B}}
\newcommand{\hC}{\hat{C}}
\newcommand{\pol}{\varepsilon}
\newcommand{\norm}[1]{\raise.3ex\hbox{:}#1\raise.3ex\hbox{:}}
\newcommand{\inn}{\!\cdot\!}
\newcommand{\bz}{\bar{z}}
\newcommand{\bw}{\bar{w}}
\newcommand{\bu}{\bar{u}}
\newcommand{\tV}{\widetilde{V}}
\newcommand{\tG}{\widetilde{G}}
\newcommand{\tB}{\widetilde{B}}
\newcommand{\tS}{\widetilde{S}}
\newcommand{\tC}{\widetilde{C}}
\newcommand{\tX}{\tilde{X}}
\newcommand{\tphi}{\tilde{\phi}}
\newcommand{\tpsi}{\tilde{\psi}}
\newcommand{\labell}[1]{\label{#1}} %\qquad_{#1}}{\label{#1}}
\newcommand{\labels}[1]{\label{#1}} %{\vskip-2ex$_{#1}%$\label{#1}} %{\label{#1}}
\newcommand{\slf}{\G} %{\hbox{{$F$}\llap{$/$}}}
\newcommand{\Tr}{{\rm Tr}}

\newcommand{\cF}{{\cal F}}
\newcommand{\cV}{{\cal V}}

\newcommand{\prt}{\partial}

\begin{document}
\begin{titlepage}
\rightline{\small hep-th/9805078 \hfill IPM/P-98/11}
\vskip 5em

\begin{center}
{\bf \huge Superstring Scattering\\[.25em]
from D-Branes Bound States}
\vskip 3em

{\large Mohammad R. Garousi\footnote{E-mail: mohammad@physics.ipm.ac.ir}}
\vskip 1em

{\em    Institute for Studies in Theoretical Physics and Mathematics IPM \\
P.O.Box 19395-5746, Tehran, Iran \\
and \\
Department of Physics, Birjand University, Birjand, Iran}
\vskip 4em

\begin{abstract}
We derive fully covariant expressions for disk scattering
amplitudes of any two massless closed strings in which mixed Neumann and
Dirichlet world-sheet boundary conditions are included.
{}From the two-point amplitudes, we derive the long range background
fields and   verify that they correspond to D$p$-brane bound state.  Also, from
the scattering amplitudes, we
calculate the linear coupling of closed string fields to D-brane
world-volume  and show that they are consistent with
Born-Infeld and Chern-Simons actions in the presence of a background field.
\end{abstract}
\end{center}
\end{titlepage}

\setcounter{footnote}{0}
\section{Introduction}

Recent exciting progress in string theory has revealed  many
new connections between superstring theories which had previously
been regarded as distinct theories\cite{conn}.
Now it appears that all string theories are
different phases of a single underlying theory, which also describes
 eleven dimensional supergravity  \cite{fmtheory}.
Within these discussions, extended
objects, other than just strings, play an important role.
Hence these developments have generated a renewed
interest in $p$-branes (\ie $p$-dimensional extended objects)
and their interactions.

In perturbative type II superstring theories, there is a remarkably simple
description
of non-perturbative $p$-branes carrying Ramond--Ramond (R-R) charges\cite{joe}.
The string background is taken to be simply flat empty space,
however interactions of closed superstrings with
these $p$-branes are described by
world-sheets with boundaries fixed to a particular surface at the
position of a $p$-brane.
The latter is accomplished by imposing Dirichlet boundary conditions
on the world-sheet fields \cite{Dbrane,Dbreview}.
Hence these objects are referred to as Dirichlet $p$-branes
(D$p$-branes) or generically as simply D-branes.
Within the type IIa theory, the D$p$-branes can have $p=0$, 2, 4, 6
or 8, while for the type IIb strings, $p$ ranges over $-1$, 1, 3, 5, 7,
 \cite{joe}.
Scattering amplitudes describing
the interactions of closed strings with D-branes was study extensively
in\cite{igora,igorb,form,ours} where 
in \cite{ours} by studying scattering amplitude of massless closed string 
from
D-branes it was verified that
the long range background 
fields around these D-branes are those of extremally charged
 $p$-brane solutions
 of the low energy effective action.

Within the massless spectrum of open string states on a D-brane
one finds a $U(1)$ gauge field \cite{leigh}.
If the D-brane carries a constant
background gauge field, this background field
 induces new couplings on the D-brane to the 
R-R form potential,
and the result may be  regarded as a bound state of D-branes\cite{douglas}. In 
scattering calculations or conformal field theory, this modifies the boundary
conditions \cite{leigh}.
In type II superstring theory, 
 interaction of closed superstrings with these D-brane bound states are
described by the same
world-sheet boundary conditions 
as for an ordinary  D-brane except for world-volume directions that carry
background  fields. In these directions
the Neumann boundary conditions are traded for mixed Neumann and Dirichlet
boundary conditions. The
interaction of massless superstrings
with  bound states of D-string and fundamental strings which is described by a  D-string with its world-volume electric field turned on,  was studied  in
\cite{gukov}. The present paper provides calculations
 on scattering amplitude of any massless closed superstring from
 bound states of two D-branes which possess a difference in dimension of
two
 using the mixed boundary condition. These bound states are described by D$p$-branes with a world-volume magnetic field turned on.

The paper is organized as follows:
In the following section we study modification of conformal field theory arising from 
mixed boundary conditions. From these study, we calculate specific $D$ and $M$
 matrices that one needs in the subsequent sections.
In section \ref{rest} we evaluate  general amplitudes for scattering any two
massless closed superstring states from a brane with arbitrary $D$ and $M$ matrices by
using previously
calculated open superstring amplitudes as was done  in \cite{ours}. Our results include the
scattering amplitudes with bosonic NS-NS and R-R states, and
also fermionic NS-R and R-NS states.
In section \ref{massless}, we examine the
massless closed string poles in the amplitudes for scattering from D-brane bound states. By comparing these
terms to those in analogous field theory calculations, we are able to
extract the long range background fields surrounding a D$p$-brane 
bound state.
Our calculations verify that these fields do correspond to those
of extremally charged $p$- and $(p-2)$-brane bound state solutions of the low
energy effective action \cite{breck}.
Next we evaluate linear coupling of closed string fields
to  the D-brane bound state  using 
the scattering amplitudes
and show that
as expected they are consistent with  Born-Infeld and Chern-Simons actions
in the presence of background gauge field. We conclude with a discussion of 
our results in Section \ref{discuss}.

\section{Conformal field theory with  background field}
\labels{conformal}

In perturbative superstring theories, to study scattering amplitude of 
some external string states 
  in conformal
field theory frame,
one usually evaluate correlation function of their corresponding vertex
operators with
use of some  standard conformal field theory 
propagators \cite{pkllsw}\footnote{I am grateful for collaborations with
Robert Myers in the results of this section.}. In trivial flat background one uses an  appropriate 
linear $\sigma$-model to
derive the propagators and define the vertex operators. While in nontrivial
background fields, one must use nonlinear $\sigma$-model to do that.  
 In
nontrivial
D-brane background the vertex operator remain unchanged whereas the standard
propagators need some modification. Alternatively, one may use a doubling
trick to convert the propagators to standard form and shift the modification
to the vertex operators\cite{ours}. In  \cite{ours} scattering amplitude of
two massless closed string states in D-brane background was studied. There
the D-brane
was assumed to be flat. In this paper we would like to turning some of
the D-brane background fields on and studying the scattering amplitude
in the resulting nontrivial background. 
We consider 
 D-branes that carry  constant gauge and/or antisymmetric Kalb-Ramond
field. The modifications arising from the appropriate  $\sigma$-model
 appear in the following boundary conditions \cite{leigh,arfaei}\footnote{
Our notation and conventions follow  those established in \cite{ours}.
So we are working on the upper-half plane 
with boundary at $y=0$ which means $\prt_y$ is
normal derivative and $\prt_x$ is tangent derivative.}:
\beqa
\prt_y X^a-i\cF^a{}_b\prt_x X^b\,\,=\,\,0&{\rm for }&a,b\,=0,1,\cdots p 
\nonumber\\
X^i\,\,=\,\,0&{\rm for}&i\,=p+1,\cdots 9
\labell{mixboundary}
\eeqa
where $\cF_{ab}$  are the  constant background fields, 
and these equations are imposed at $y=0$.
In general one may choose to have non-zero
background $\cF_{ab}$ in all direction of D-brane  but for simplicity we are
interested in adding a constant background field in only two spatial 
direction of the
D-brane world-volume \eg $\cF_{12}=-\cF_{21}=\cF$. In this case
the boundary condition (\ref{mixboundary}) becomes
\beqa
\prt_y X^c&=&0  \nonumber\\
\prt_y X^1-i\cF\prt_x X^2&=&0 \nonumber\\
\prt_y X^2+i\cF\prt_x X^1&=&0  \nonumber\\
X^i&=&0
\labell{mixone}
\eeqa
where $c=0,3,4,\cdots p$.
For special case of $\cF=0$, one has the Neumann boundary condition on
all $p$ dimensions which describe D$p$-brane. 
The Dirichlet boundary condition  on the remaining coordinates
fixes the position of the D$p$-brane at $X^i=0$. While for large $\cF$, the 
Neumann boundary condition are imposed only on $p-2$ dimensions which
describe D$(p-2)$-brane. Here Dirichlet-like boundary conditions are imposed
on two dimensions, \ie $\prt_xX^{1,2}=0$, which does not fix position of
the D$(p-2)$-brane in these dimensions. For arbitrary value of $\cF$ one
has both D$p$-brane and D$(p-2)$-brane which the latter delocalized on the
former, and the result may be considered as bound state of D$p$- and D$(p-2)$-branes.

Now we have to understand the modification of the conformal field theory
propagators arising from these mixed boundary conditions. To this end consider
the following general expression for propagator of $X^{\mu}(z,\bz)$ fields:
\beqa
<X^{\mu}(z,\bz)\,X^{\nu}(w,\bw)>
&=&-\eta^{\mu\nu}\log(z-w)-\eta^{\mu\nu}\log(\bz-\bw) \nonumber\\
&&-D^{\mu\nu}\log(z-\bw)-D^{\nu\mu}\log(\bz-w)
\labell{pro1}
\eeqa
where $D^{\mu\nu}$ is a constant matrix that has to be fixed by imposing
appropriate
boundary condition. 
When both $z$ and $w$ are on the boundary of the world-sheet, 
\ie $z=\bz\equiv x_1$
and $w=\bw\equiv x_2$, the propagator (\ref{pro1}) becomes
\beqa
<X^{\mu}(x_1,x_1)\,X^{\nu}(x_2,x_2)>&=&
-2(\eta^{\mu\nu}+(D^S)^{\mu\nu})\log(x_1-x_2)
+i\pi(D^A)^{\mu\nu}\Theta(x_1-x_2)\nonumber
\labell{pro11}
\eeqa
where $D^S(D^A)$ is symmetric(antisymmetric) part of the $D$ matrix, and
$\Theta(x_1-x_2)=1(-1)$ if $x_1>x_2(x_1<x_2)$. Note that the last term in above equation
which stems from the cut line of the Log function 
is zero for the systems that their $D^{\mu\nu}$ matrix is 
symmetric.
Now the $D$ matrix can be found for different boundary conditions.
For the case that there is no boundary one finds
 $D^{\mu\nu}=0$, 
the standard propagators. In  \cite{ours} the matrix $D^{\mu\nu}$
for flat D-brane
 was found to be $D_0^{\mu\nu}=V_0^{\mu\nu}+N^{\mu\nu}$
where $N^{\mu\nu}(V_0^{\mu\nu})$ projects vectors into subspace
orthogonal(parallel) to D-brane. Using
boundary conditions (\ref{mixone}),
 one  finds that the matrix $D^{\mu\nu}$
for the D-brane bound state is the same as $D_0^{\mu\nu}$
except in $X^1$ and $X^2$ directions, that is
\[
D^{\mu}{}_{\nu}\,=\,{D_0}^{\mu}{}_{\lambda}R^{\lambda}{}_{\nu}\,
=\,R^{\mu}{}_{\lambda}
{D_0}^{\lambda}{}_{\nu}
\]
where
\[
{R}=\pmatrix{\cos 2\t & -\sin 2\t\cr
\sin 2\t & \cos 2\t\cr}
\]
is rotation matrix  for angle $2\t$. Here we defined $\cos 2\t\equiv
(1-\cF^2)/(1+\cF^2)$ and $\sin 2\t\equiv 2\cF/(1+\cF^2)$. 
Note that this $D$ matrix satisfies
 $D^{\mu}{}_{\alpha}D^{\alpha\nu}=\eta^{\mu\nu}$.
Now writing
$X^{\mu}(z,\bz)=X^{\mu}(z)+\tX^{\mu}(\bz)$, eq.~(\ref{pro1}) yields
\beqa
<X^{\mu}(z)\,X^{\nu}(w)>&=&-\eta^{\mu\nu}\log(z-w) \nonumber\\
<\tX^{\mu}(\bz)\,\tX^{\nu}(\bw)>&=&-\eta^{\mu\nu}\log(\bz-\bw)  \nonumber\\
<X^{\mu}(z)\,\tX^{\nu}(\bw)>&=&-D^{\mu\nu}\log(z-\bw)
\labell{pro2}\nonumber
\eeqa
Similarly, propagator of $\psi^{\mu}$ fields, world-sheet super partner
of $X^{\mu}$, are modified as
\beqa
<\psi^{\mu}(z)\,\psi^{\nu}(w)>&=&-\frac{\eta^{\mu\nu}}{z-w} \nonumber\\
<\tpsi^{\mu}(\bz)\,\tpsi^{\nu}(\bw)>&=&-\frac{\eta^{\mu\nu}}{\bz-\bw} 
\nonumber\\
<\psi^{\mu}(z)\,\tpsi^{\nu}(\bw)>&=&-\frac{D^{\mu\nu}}{z-\bw}
\labell{pro3}\nonumber
\eeqa
and propagator of ghost fields remain unchanged and are
\beqa
<\phi(z)\,\phi(w)>&=&-\log(z-w) \nonumber \\
<\phi(z)\,\tphi(\bw)>&=&-\log(z-\bw)\nonumber \\
<\tphi(\bz)\,\tphi(\bw)>&=&-\log(\bz-\bw)
\labell{pro4}\nonumber
\eeqa
These propagators
  can be transformed to the standard
form by the doubling trick
\beqa
\tX^{\mu}(\bz)\longrightarrow D^{\mu}{}_{\nu}X^{\nu}(\bz) &\,\,\,\,\,
\tpsi^{\mu}(\bz)\longrightarrow D^{\mu}{}_{\nu}\psi^{\nu}(\bz)&\,\,\,\,\,
\tphi(\bz)\longrightarrow \phi (\bz)
\labell{trick}
\eeqa
These replacements in effect extend the left-moving 
fields to the entire complex plane and
shift modification arising from mixed boundary condition
from propagators to vertex operators. Under replacements (\ref{trick}), 
propagator of the left-moving fields on the boundary become
\beqa
<X^{\mu}(x_1)\,X^{\nu}(x_2)>&=&-\eta^{\mu\nu}\log(x_1-x_2)-i\frac{\pi}{2}
\cF^{\mu\nu}\Theta(x_1-x_2)\nonumber\\
<\psi^{\mu}(x_1)\,\psi^{\nu}(x_2)>&=&-\eta^{\mu\nu}\log(x_1-x_2)\nonumber\\
<\phi(x_1)\,\phi(x_2)>&=&-\log(x_1-x_2)\,\,\, .\nonumber
\eeqa
In scattering amplitude of boundary states, effect of the last term in 
the first line above is to add some phase factor to the 
scattering amplitude \cite{gukov}.

In order for studying scattering amplitude of
all closed string states from D-brane with background field, one has to also
know
how to use the doubling trick for spin operator. Right-moving spin operator is
replaced by
\beqa
\tS_A(\bz)\longrightarrow M_A{}^B S_B(\bz)
\labell{trick2}
\eeqa
where the constant $M$ matrix is defined by\cite{ours}
\beqa
\g^{\mu}&=&(D^{-1})^{\mu}{}_{\nu}M^{-1}\g^{\nu}M 
\labell{Mmatrix}
\eeqa
Now using the decomposition of $D=D_0R$, one should be able to decompose
$M=\O(R)M_0$ where $M_0$ is the standard $M$ matrix which was derived in
\cite{ours} (for D-brane with $\cF=0$),
\beq
M^p_0=\left\lbrace\matrix{
&\frac{\pm i}{(p+1)!}\,(\e^v)_{\mu_0\cdots\mu_p}\,\g^{\mu_0}\cdots\g^{\mu_p}
\,\hphantom{\g_{11}}&\qquad {\rm for}\ p+1\ {\rm odd}\cr
&\frac{\pm 1}{(p+1)!}\,(\e^v)_{\mu_0\cdots\mu_p}\,\g^{\mu_0}\cdots\g^{\mu_p}
\,\g_{11}&\qquad {\rm for}\ p+1\ {\rm even}\ .\cr}
\right.
\labell{finalm}
\eeq
where $\e^v$ is the world-volume form of D-brane, 
and $\O(R)$ is the spinor rotation matrix for
the angle $2\t$. (Note that since $D_0$ and $R$ commute, the order of $\O$
and $M_0$ in $M$ should not be important.) To see this, using the 
properties of $M_0$ matrix, $\g^dM_0=M_0\g^d$,
one can write (\ref{Mmatrix}) as
\beqa
\O^{-1}\g^d\O&=&D^d{}_e\g^e\,\,=\,\,R^d{}_e\g^e
\labell{Omatrix}
\eeqa
where $d,e=1,2$. In component form it is
\beqa
\O^{-1}\g^1\O&=&\g^1\cos 2\t-\g^2\sin 2\t  \nonumber\\
\O^{-1}\g^2\O&=&\g^1\sin 2\t+\g^2\cos 2\t
\nonumber
\eeqa
These equations can be solved for $\O$ matrix by using Baker-Hausdorff
expansion
and the result is
\[
\O\,\,=\,\, exp[\t\g^1\g^2]\,\,=\,\,\cos\t+\g^1\g^2\sin\t
\]
Since $M_0\g^d\g^e=\g^d\g^e M_0$, one realizes that $M_0\O=\O M_0$ as
anticipated above.
Therefore $M$ matrix for D-brane with constant background $\cF$ field is
\beqa
M^p&=&M_0^p\cos\t-M_0^{p-2}\sin\t
\labell{Mmatrix1}\nonumber
\eeqa
This form for M matrix was also found in \cite{vecchia}.
Some trigonometric identities can be used to show
\beqa
\cos\t=\frac{1}{\sqrt{1+\cF^2}}&\,,\,& \sin\t=\frac{\cF}{\sqrt{1+\cF^2}}
\labell{cossin}
\eeqa
Now that $D$ and $M$ matrices have been found, one may use the replacement
(\ref{trick}) and
(\ref{trick2}) to convert evaluation of scattering amplitude to standard
conformal
field theory calculation.

\section{Two-point amplitudes} \labels{rest}

A systematic approach for studying scattering amplitude of all massless closed
superstring
states from D-brane was developed in \cite{ours} where it was shown 
that there is
a direct relation between four-point amplitude of type I theory and two-point
amplitude
of type II theory in a D-brane background.
Since the former amplitudes are well-known for all 
massless states, a simple
transformation involving among other things $D_0$ and $M_0$ matrices gives the
two-point amplitudes.
The difference between scattering
amplitude of massless closed string states from D-brane with $\cF=0$ and
$\cF\neq 0$
is in the specific form of $D$ and $M$ matrices, hence one may use the former results to evaluate scattering amplitude of the latter. 

General form of  amplitudes for   
scattering any two massless closed superstring states 
from a brane with unspecified $D$ and $M$ matrices was given in \cite{ours}.
However, there we implicitly assumed that the $D$ matrix being   symmetric.
We should release that assumption in order to include D-branes with constant
background fields. Therefore, we redo those calculations with the most general form for $D$ and $M$ matrices.
So we begin with the well-known four-particle open superstring amplitudes involving massless vectors as well as spinors which may be expressed in the 
form \cite{jhsreport}\footnote{Here and in the subsequent amplitudes, we
omit the Dirac delta-function which imposes momentum conservation.}
\beq
A(1,\,2,\,3,\,4)=-\frac{1}{2}g^2
\frac{\G(4k_1\inn k_2)\G(4k_1\inn k_4)}{\G(1+4k_1\inn k_2+4k_1\inn k_4)}
K(1,\,2,\,3,\,4)\ \ .
\labell{opentot}
\eeq
The various kinematic factors are then given by
\beqa
K(\z_1\,\z_2\,\z_3\,\z_4)&=&-16k_2\inn k_3 k_2\inn
k_4\,\z_1\inn\z_2\z_3\inn\z_4
-16k_1\inn k_2\left(\z_1\inn k_4\z_3\inn k_2\z_2\inn\z_4 \right.
\labell{openzero} \\
&&\left.\,+\z_2\inn k_3\z_4\inn k_1\z_1\inn\z_3+\z_1\inn k_3\z_4\inn
k_2\z_2\inn\z_3
+\z_2\inn k_4\z_3\inn k_1\z_1\inn\z_4\right)\nonumber \\
&&\,\,\Big\{1,2,3,4\longrightarrow 1,3,2,4\Big\}+\Big\{1,2,3,4\longrightarrow
1,4,3,2\Big\}
\nonumber \\
K(u_1,\,u_2,\,u_3,\,u_4)&=&
-2\,k_1\inn k_2\,\bu_2 \g^\mu u_3\,\bu_1\g_\mu u_4
+2\,k_1\inn k_4\,\bu_1\g^\mu u_2\,\bu_4\g_\mu u_3
\labell{openone}\\
K(u_1,\,\z_2,\,\z_3,\,u_4)&=&
2i\sqrt{2}\,k_1\inn k_4\,\bu_1\g\inn \z_2\g\inn(k_3+k_4)\g\inn\z_3 u_4
\labell{opentwo}\\
&&-4i\sqrt{2}\,k_1\inn k_2\,\left(\bu_1\g\inn\z_3u_4\,k_3\inn\z_2
-\bu_1\g\inn\z_2u_4\,k_2\inn\z_3-\bu_1\g\inn k_3u_4\,\z_2\inn\z_3\right)
\nonumber\\
K(u_1,\,\z_2,\,u_3,\,\z_4)&=&
-2i\sqrt{2}\,k_1\inn k_4\,\bu_1\g\inn \z_2\g\inn(k_3+k_4)\g\inn\z_4 u_3
\labell{openthree}\\
&&\qquad\quad\qquad-2i\sqrt{2}\,k_1\inn k_2\,\bu_1\g\inn
\z_4\g\inn(k_2+k_3)\g\inn\z_2 u_3
\ \ .
\nonumber
\eeqa
In translating these results to the  two-point amplitudes in D-brane bound
state background, schematically
ones associates (some cyclic permutation of) $(1,2,3,4)$ in the open string
amplitude with $(1_L,2_L,2_R,$ $1_R)$ in the closed string amplitude,
where here the subscripts $L$ and $R$ denote the left- and right-moving
components of the closed string states. 
An appropriate transformation of momenta
and polarization tensors from open  to closed superstring
states gives scattering amplitude of closed superstring states from the
 D-brane
bound states.
The D-brane scattering amplitudes then take a universal form
\beq
A(1,\,2)=-i\,\frac{\kappa\,T_p\sqrt{1+\cF^2}}{2}\,
\frac{\G(-t/2)\G(2q^2)}{\G(1-t/2+2q^2)}
K(1,\,2)\,\,.
\labell{ampuniverse}
\eeq
Here we normalized the amplitude by replacing 
$g^2\longrightarrow i\kappa^(2-\chi)
T_{p,p-2}$
where the Euler number $\chi=1$ for disk and $\kappa$ and $T_{p,p-2}=T_p\sqrt{1+\cF^2}$ are the closed string and  the D-brane 
coupling
constants,
respectively
\cite{gukov}. 
For later discussions, it
is
useful to divide the kinematic factor as
\beq
K(1,\,2)=2q^2\,a_1(1,\,2)+\frac{t}{2}\,a_2(1,\,2)
\labell{subdivide}
\eeq
Then $a_1(1,\,2)$ will be essentially
the residue of the massless $t$-channel pole, which will become
important for the analysis in sect.~\ref{massless}.
Now, it simply remains to translate
the kinematic factors (\ref{openzero})-(\ref{openthree}) in the
appropriate way.

\subsection{NS-NS scattering amplitudes}

We begin by calculating the amplitudes describing the
scattering of two massless \hbox{NS-NS} states from a Dirichlet $p$-brane
bound state
(\ie the scattering of gravitons, dilatons
or Kalb-Ramond (antisymmetric tensor) states).
The amplitudes are calculated as two closed string vertex operator insertions
on a disk with  boundary conditions
(\ref{mixone}).
The amplitude may  be written as
\beq
A\simeq\int d^2\!z_1\, d^2\!z_2\ \langle\, V_1(z_1,\bz_1)
\ V_2(z_2,\bz_2)\,\rangle
\labell{ampone}
\eeq
where the vertex operators are
\beqa
V_1(z_1,\bz_1)&=&\pol_{1\mu\nu}\,\norm{V_{-1}^\mu(p_1,z_1)}
\ \norm{\tV_{-1}^\nu(p_1,\bz_1)}
\nonumber\\
V_2(z_2,\bz_2)&=&\pol_{2\mu\nu}\,\norm{V_{0}^\mu(p_2,z_2)}
\ \norm{\tV_{0}^\nu(p_2,\bz_2)}\ \ .
\labell{vertone}
\eeqa
The holomorphic components above are given by
\beqa
V_{-1}^\mu(p_1,z_1)&=&e^{-\phi(z_1)}\,\psi^{\mu}(z_1)\,e^{ip_1\cdot X(z_1)}
\nonumber\\
V_0^\mu(p_2,z_2)&=&\left(\partial X^\mu(z_2)+ip_2\inn \psi(z_2)\psi^{\mu}(z_2)
\right)\,e^{ip_2\cdot X(z_2)}\ \ .
\labell{vertright}
\eeqa
The anti-holomorphic components take the same form as in eq.~(\ref{vertright})
but with the left-moving fields replaced by their
right-moving counterparts -- \ie
$X(z)\rightarrow\tX(\bz)$, $\psi(z)\rightarrow\tpsi(\bz)$, and
$\phi(z)\rightarrow\tphi(\bz)$. As usual, the momenta and polarization
tensors satisfy
\[
p_i^2=0\ ,\qquad\qquad p_i^\mu\,\pol_{i\mu\nu}=0=\pol_{i\mu\nu}\,p_i^\nu\ \ 
\]
and the various physical states would be represented with
\beqa
{\rm graviton}:&&\quad \pol_{i\mu\nu}=\pol_{i\nu\mu},\ \ \pol_{i\mu}{}^\mu=0
\nonumber\\
{\rm dilaton}:&&\quad \pol_{i\mu\nu}=\frac{1}{\sqrt{8}}\left(\eta_{\mu\nu}
-p_{i\mu}\ell_{i\nu}-\ell_{i\mu}p_{i\nu}\right)
\ \ {\rm where}\ p_i\inn\ell_i=1
\nonumber\\
{\rm Kalb-Ramond}:&&\quad \pol_{i\mu\nu}=-\pol_{i\nu\mu}\ \ .
\labell{nspol}
\eeqa
In order for dealing with standard conformal field theory propagators, we use
the doubling trick (\ref{trick}) which convert
the vertex operators (\ref{vertone}) to
\beqar
V_1(z_1,\bz_1)&=&\pol_{1\mu\lambda}D^\lambda{}_\nu\,\norm{V_{-1}^\mu(p_1,z_1)}
\ \norm{V_{-1}^\nu(D^T\inn p_1,\bz_1)}
\\
V_2(z_2,\bz_2)&=&\pol_{2\mu\lambda}D^\lambda{}_\nu\,\norm{V_{0}^\mu(p_2,z_2)}
\ \norm{V_{0}^\nu(D^T\inn p_2,\bz_2)}
\eeqar
using only the expressions in eq.~(\ref{vertright}).

Now following \cite{ours}, one finds the final result of scattering amplitude
(\ref{ampuniverse}) by replacing
\beqa
2k_1^\mu\rightarrow p_1^\mu&&2k_4^\mu\rightarrow (D^T\inn p_1)^\mu
\nonumber\\
2k_2^\mu\rightarrow p_2^\mu&&2k_3^\mu\rightarrow (D^T\inn p_2)^\mu
\nonumber\\
\z_{1\mu}\,\otimes\,\z_{4\nu}&\rightarrow \pol_{1\mu\lambda}D^\lambda{}_\nu&
\nonumber\\
\z_{2\mu}\,\otimes\,\z_{3\nu}&\rightarrow \pol_{2\mu\lambda}D^\lambda{}_\nu
& ,
\labell{transone}
\eeqa
in eqs.~(\ref{opentot}) and (\ref{openzero}).
Hence,
the final result may be written as
\beq
A=-i\frac{\kappa\,T_p\sqrt{1+\cF^2}}{2}\,
\frac{\G(-t/2)\G(2q^2)}{\G(1-t/2+2q^2)}
\left(2q^2\,a_1+\frac{t}{2}\,a_2\right)
\labell{finone}
\eeq
where $t=-2p_1\inn p_2$ is the momentum transfer to the brane, 
 and $q^2={1\over2}p_1\inn D\inn p_1$ is the momentum flowing parallel to
the world-volume of the brane. 
 The kinematic factors above are:
\beqa
a_1&=&{\rm Tr}(\pol_1\inn D)\,p_1\inn \pol_2 \inn p_1 -p_1\inn\pol_2\inn
D\inn\pol_1\inn p_2 - p_1\inn\pol_2\inn\pol_1^T \inn D^T\inn p_1
-p_1\inn\pol^T_2\inn\pol_1\inn D\inn p_1
\nonumber\\
&&-\frac{1}{2}(p_1\inn\pol_2\inn \pol^T_1\inn p_2+p_2\inn\pol^T_1\inn\pol_2
\inn p_1)+
q^2\,{\rm Tr}(\pol_1\inn\pol_2^T)
+\Big\{1\longleftrightarrow 2\Big\}
\labell{fintwo}\\
\nonumber\\
a_2&=&{\rm Tr}(\pol_1\inn D)\,(p_1\inn\pol_2\inn D\inn p_2 + p_2\inn
D\inn\pol_2\inn p_1 +p_2\inn D\inn\pol_2\inn D\inn p_2)
+p_1\inn D\inn\pol_1\inn D\inn\pol_2\inn D\inn p_2
\nonumber\\
&&-\frac{1}{2}(p_2\inn
D\inn\pol_2\inn\pol_1^T\inn D^T\inn p_1+p_1\inn D^T\inn\pol^T_1\inn\pol_2\inn
D\inn p_2)
+q^2\,{\rm Tr}(\pol_1\inn D\inn \pol_2\inn D)
\nonumber\\
&&-q^2\,{\rm Tr}(\pol_1\inn\pol_2^T)
-{\rm Tr}(\pol_1\inn D) {\rm Tr}(\pol_2\inn D)\,(q^2-t/4)
+\Big\{1\longleftrightarrow 2 \Big\}\ \ .
\labell{finthree}\nonumber
\eeqa
Our notation is such that \eg $p_1\inn\pol_2\inn\pol_1^T \inn D^T\inn p_1
=p_1^\mu\,\pol_{2\mu\nu}\,
\pol_1{}^{\lambda\nu}\,(D^T)_{\lambda\rho}\,p_1^\rho$.
Conservation of  momentum under replacement (\ref{transone}) becomes
\beqa
(p_1+p_1\inn D+p_2+p_2\inn D)^{\mu}&=&0 
\label{momentum}
\eeqa
Now for the Dirichlet directions (\ie those orthogonal to the D-brane), the
left hand-side automatically vanishes and so there
is no momentum conservation in those directions. For the world-volume
directions
other than $\mu=1,2$, this equation yields as before that $(p_1+p_2)^{\mu}=0$.
The
equation for $\mu=1,2$ are more complicated, but also require that momentum is
conserved
in those directions. Thus just as in the case with $\cF=0$, momentum is
conserved in the
world-volume directions, \ie $(p_1+p_2)\cdot V_0^{\mu}=0$.

\subsection{R-R boson amplitude}

The next simple case is using eqs.~(\ref{opentot}) and (\ref{openone}) to
calculate the amplitude describing two R-R states scattering from the
D-brane bound state. The latter amplitude would be written as
\[
A\simeq \int d^2\!z_1\, d^2\!z_2\ \langle\, V_1(z_1,\bz_1)
\ V_2(z_2,\bz_2)\,\rangle
\]
where the vertex operators are
\beq
V_i(z_i,\bz_i)=(P_-\,\slf_{i(n)})^{AB}\,\norm{V_{-1/2\,A}(p_i,z_i)}
\ \norm{\tV_{-1/2\,B}(p_i,\bz_i)}\ \ .
\labell{vertspin}
\eeq
The holomorphic components above are given by
\beq
V_{-1/2\,A}(p_i,z_i)=e^{-\phi(z_i)/2}\,S_A(z_i)\,e^{ip_i\cdot X(z_i)}
\labell{vertspinor}
\eeq
and the anti-holomorphic components have the same form, but with the
left-moving fields replaced by their right-moving counterparts. As before
we use eq.~(\ref{trick}) to replace anti-holomorphic components $\tX^\mu$ and
$\tphi$ in $\tV_{-1/2\,B}$
by their corresponding holomorphic components.
Similarly, the right-moving spin field is replaced using (\ref{trick2}) by
its corresponding left-moving spin field.
With  replacement (\ref{trick2}), only standard correlators of the
spin fields \cite{danf,pkllsw,fms} appear in the subsequent calculations.
We have explicitly included the chiral projection operator
$P_-=(1-\g_{11})/2$ in vertex operator (\ref{vertspin}),
so that our calculations are always made with the full 32$\times$32 Dirac
matrices of ten dimensions. We have also defined
\beq
\slf_{i(n)}=\frac{a_n}{n!}F^i_{\mu_1\cdots\mu_n}\,\g^{\mu_1}\cdots
\g^{\mu_n}\ \
\labell{self}
\eeq
where $a_n=i$ for
the $n=2$ and 4 fields in the type IIa theory, while
$a_n=1$ for $n=1$, 3 and 5 in the type IIb theory. In eq.~(\ref{self}),
$F^i_{\mu_1\cdots\mu_n}$ is the linearized $n$-form field strength
with
\beqa
F^i_{\mu_1\cdots\mu_n}&=&i\, n\, p_{i[\mu_1}\pol_{i\mu_2\cdots\mu_n]}
\labell{strong}\nonumber\\
&=&i\,p_{i\mu_1}\pol_{i\mu_2\cdots\mu_n} \pm\ {\rm cyclic\ permutations}
\nonumber
\eeqa
where $p_i^2=0$ and $p_i^\mu\,\pol_{i\mu\mu_3\cdots\mu_n}=0$.
Hence the appropriate substitutions for the open string
amplitude (\ref{openone}) to derive the  D-brane  amplitude  are
\beqa
2k_1^\mu\rightarrow p_1^\mu&&2k_4^\mu\rightarrow (D^T\inn p_1)^\mu
\nonumber\\
2k_2^\mu\rightarrow p_2^\mu&&2k_3^\mu\rightarrow (D^T\inn p_2)^\mu
\nonumber\\
u_{1A}\,\otimes\,u_{4B}&\rightarrow (P_-\,\slf_{1(n)}M)_{AB}&
\nonumber\\
u_{2A}\,\otimes\,u_{3B}&\rightarrow (P_-\,\slf_{2(m)}M)_{AB}
& .
\labell{transtwo}
\eeqa
The resulting kinematic factor is:
\beqa
a_1^{R-R,R-R}&=&
-\frac{1}{2}\Tr(P_-\slf_{1(n)}M\g_\mu C^{-1}
M^T\slf^T_{2(m)}C\g^\mu)\labell{kinrrrr}\nonumber
\\ \nonumber\\
a_2^{R-R,R-R}&=&\frac{1}{2}\Tr(P_-\slf_{1(n)}M\g_\mu)\,\Tr(P_-\slf_{2(m)}
M\g^\mu)
\labell{a2rrrr}\nonumber
\eeqa
where the trace is over the $32\times 32$ Dirac matrices of ten dimensions.
Performing
these traces, one would find the kinematic factor in terms of only momenta and
polarization
tensors.

\subsection{NS-NS and R-R amplitude}

The next case is calculating the amplitude describing one R-R and
one NS-NS state scattering from a Dirichlet brane with non-zero $\cF$
using eqs.~(\ref{opentot}) and (\ref{opentwo}).
The appropriate substitutions to derive the Dirichlet amplitude are already
derived for the previous amplitudes in eqs.~(\ref{transone})
and (\ref{transtwo})
\beqar
2k_1^\mu\rightarrow p_1^\mu&&2k_4^\mu\rightarrow (D^T\inn p_1)^\mu
\\
2k_2^\mu\rightarrow p_2^\mu&&2k_3^\mu\rightarrow (D^T\inn p_2)^\mu
\nonumber\\
u_{1A}\,\otimes\,u_{4B}&\rightarrow (P_-\,\slf_{1(n)}M)_{AB}&
\\
\ \z_{2\mu}\,\otimes\,\z_{3\nu}\hphantom{B}
&\rightarrow \pol_{2\mu\lambda}D^\lambda{}_\nu\ \
\hphantom{M)_{AB}}&\ \ .
\eeqar
The resulting kinematic factor is then:
\beqa
a_1^{R-R,NS-NS}&=&
\frac{i}{2\sqrt{2}}\Tr[P_-\slf_{1(n)}M\g^\nu\g\inn(p_1+p_2)\g^\mu]\,(
\pol_2\cdot D)_{\mu\nu}
\labell{kinrrnn} \\
\nonumber\\
a_2^{R-R,NS-NS}&=& -\frac{2i}{\sqrt{2}}\left[\Tr(P_-\slf_{1(n)}M\g\inn
D^T\inn\pol^T_2\inn D^T\inn p_2)
-\Tr(P_-\slf_{1(n)}M\g\inn \pol_2\inn D\inn p_2)\right.
\nonumber\\
&&\left.\ \
\qquad\qquad-\Tr(P_-\slf_{1(n)}M\g\inn D^T\inn p_2)\,\Tr(\pol_2\inn
D)\right]\,\,.
\labell{kinrrnn2}\nonumber
\eeqa
This amplitude will be of particular interest
in the following section for determining the background R-R fields
in sect.~\ref{long}.
\subsection{Fermion amplitudes}

Following \cite{ours}, the kinematic factor for scattering of two 
fermions can also
be
evaluated and the result would be those in  \cite{ours} with some of the $D$
matrices replaced
by $D^T$, that is
\beqa
K^{R-NS,R-NS}&=&
i\sqrt{2}\,q^2(\psi_{2}\cdot p_{1}\g\inn D^T\inn P_-\psi_{1}-\psi_{2}\cdot
D\inn \g\, p_{2}\inn P_-\psi_{1}
\nonumber\\
&&\qquad\qquad-{\psi_{2}}^{\mu}\,\g\inn D^T \inn p_{1}\,P_-\psi_{1\mu})
\nonumber\\
&&\  \ +i\frac{t}{4\sqrt{2}}\left(\psi_{2}\cdot D\inn\g\,\g\inn(p_{1}+D^T\inn
p_{1})\,\g\inn D^T\inn P_-\psi_{1}\right)\nonumber\\
\nonumber \\
K^{NS-R,NS-R}&=&
i\sqrt{2}\,q^2(\psi_1\inn p_2\,M^{-1}\,\g\inn
M\,P_\pm\psi_1-\psi_2\,M^{-1}\inn\g\, M\,P_{\pm}\,p_1\inn\psi_2
\nonumber\\
&&\qquad\qquad-{\psi_1}^{\mu}\,M^{-1}\,\g\inn p_2\,M\,P_\pm\psi_{2\mu})
\nonumber\\
&&\ \ +i\frac{t}{4\sqrt{2}}(\psi_1\,M^{-1}\inn\g\,\g\inn (p_2+D^T\inn p_2)
\,\g\inn M\,P_\pm\psi_2)\nonumber\\
\nonumber\\
K^{R-NS,NS-R}&=&
-i\frac{q^2}{\sqrt{2}}(\psi_1\,P_+\cdot
D_{\nu}\,\g^{\mu}\,\g\inn(p_1+p_2)\,\g^{\nu}\,M\,{\psi_2}_{\mu})
\nonumber\\
&&-i\frac{t}{4\sqrt{2}}(\psi_1\,P_+\cdot D\inn\g\,\g\inn(p_2+D^T\inn p_2)
\,\g\inn M\,\psi_2)\nonumber
\eeqa
where $\psi_1$ and $\psi_2$ are spinor polarizations. 

Up to here we never used properties of $D$ and $M$ matrices in evaluating these boson and 
fermion amplitudes, hence they satisfy  for general $D$ and $M$ matrices. One    gets various
different results correspond to choosing an explicit $D$ and $M$ matrices and
explicitly evaluating the amplitudes. For example, one may use these amplitudes
to evaluate amplitudes for scattering any two massless string states 
from D-branes with background fields in more than one plane or the D-branes
that appear in the Appendix. In the rest of this paper, we specify the
$D$ and $M$ matrices to those that correspond to the D-branes with constant
background field $\cF$.

\section{Massless $t$-channel poles} \labels{massless}

Recall that in the scattering amplitudes, the momentum transfer
to the D-brane bound state is $t=-(p_1+p_2)^2$. Given the general form
of the string amplitudes in eqs.~(\ref{ampuniverse}-\ref{subdivide}),
one can expand these amplitudes as an infinite sum of terms reflecting
the infinite tower of closed string states that couple to the D-brane in
the $t$-channel\footnote{We explicitly restore $\alpha'$ here.
Otherwise our conventions set $\alpha'=2$.}
 (\ie terms with poles at $\alpha'\,t=\alpha'\,m^2=4n$
with $n=0,1,2,\ldots$).
For low momentum transfer,
\ie $\alpha'\,t<<1$, the first term representing the exchange of
massless string states dominates. In this case, eqs.~(\ref{ampuniverse}
-- \ref{subdivide}) reduce to
\beq
A\simeq i\,{\kappa\, T_p\sqrt{1+\cF^2}}\, \frac{a_1}{t}\ \ .
\labell{pole}
\eeq
One can reproduce these long-range interactions with a calculation
in the low energy effective field theory in which D$p$-brane source
terms are added to the field theory action.

\subsection{NS-NS sources} \labels{nsnsns}

The NS-NS sector is common to both type II
superstring theories, and so the same low energy effective action
describes the graviton, dilaton and Kalb-Ramond fields in both
theories. The latter may be written as
\beq
I^{NS-NS}=
\int d^{10}\!x\,\sqrt{-g}\,\left [\frac{1}{2\ka^2}R
-\frac{1}{2}(\na \phi)^2-\frac{1}{6}\ H^2\,e^{-{\sqrt{2}\ka}\phi}
\right]
\labell{nsaction}
\eeq
where $H_{\alpha\mu\nu} =\partial_\alpha B_{\mu\nu}+\partial_\nu
B_{\alpha \mu}+\partial_\mu B_{\nu \alpha}$.
Given this low energy effective action, one can calculate the
different propagators, interactions, and subsequently scattering amplitudes
for these three massless NS-NS particles. In doing so, one
defines the graviton field by $g_{\mu\nu}=\eta_{\mu\nu}+2\ka\,h_{\mu\nu}$.

To consider the scattering of these particles from a D$p$-brane bound state, we would
supplement the low energy action with source terms for the brane as
follows:
\beq
I^{NS-NS}_{source}=\int d^{10}\!x\,\left[S_B^{\mu\nu}\,B_{\mu\nu}+S_\phi\,
\phi+ S_h^{\mu\nu}\,h_{\mu\nu}\right]\ \ .
\labell{souraction}
\eeq
Note that at least to leading order,
$S_B$, $S_\phi$ and $S_h$ above will be $\delta$-function sources which are
only non-vanishing at $x^i=0$ where $x^i$ is position of D-brane.
%\begin{figure}
%\centerline{\epsfxsize 2.4 truein \epsfbox {p1.eps}}
%\caption{Feynman diagram for graviton-dilaton scattering from a D-brane.}
%\labels{phih}
%\end{figure}
We begin by determining the dilaton source $S_\phi$. To this end, we consider
a scattering process in which an external dilaton is converted to
a graviton.
Examining the low energy action (\ref{nsaction}), one finds that
the only relevant three-point interaction is one graviton coupling
to two dilatons through the dilaton kinetic term. Thus,
the only particle appearing in the $t$-channel is the dilaton, and
hence this amplitude will uniquely determine $S_\phi$.
The field theory amplitude may be written
\[
A'_{h\phi}=i\tilde{S}_\phi(k)\,\tG_{\phi}(k^2)\,\tV_{h\phi \phi}
(\pol_1,p_1,p_2)
\]
where $\tilde{S}_\phi(k)$ is the Fourier transform of the dilaton source,
\beqa
\tG_\phi(k^2)&=&-i/k^2
\labell{dprop}
\eeqa
is the dilaton's Feynman propagator, and
\[
\tV_{h\phi \phi}=-i\,2\ka\, p_2\inn \pol_1\inn k
\]
is the vertex factor for the graviton-dilaton-dilaton interaction.
Here, $\pol_1$ is the graviton polarization tensor, and
$k^\mu=-(p_1+p_2)^\mu$ is the $t$-channel momentum. (We have
not included in $A'_{h\phi}$ the Dirac delta-function which imposes
momentum conservation in the directions parallel to the D$p$-brane bound state.)
The analogous  string amplitude $A^{\phi h}$ is constructed
from eq.~(\ref{finone}) by inserting the appropriate
external polarization tensors from eq.~(\ref{nspol}).
To compare $A'_{h\phi}$
with the massless $t$-channel pole in  $A^{\phi h}$,
one needs to evaluate $a_1$ for graviton and dilaton. Inserting the
gravitation and dilaton
polarization tensors from eq.~(\ref{nspol}), we  find
\beqa
a_1^{\phi h}&=&\frac{2+\Tr(D)}{2\sqrt{2}}\,p_2\cdot \pol_1\cdot p_2+\cdots
\labell{DG}\nonumber
\eeqa
where the dots represent the terms which are either proportional to $k^2$ 
and hence do not appear as simple pole or
 canceled against
$a_2^{\phi h}$ term in the whole string amplitude (\ref{finone}). Now
comparing $A'_{h\phi}$ with (\ref{pole}), one finds
agreement by setting
\beq
\tS_\phi(k)=
-\frac{T_p\sqrt{1+\cF^2}}{4\sqrt{2}}\left (2+\Tr(D)\right)=-
\frac{T_p\sqrt{1+\cF^2}}{2\sqrt{2}}(\cos(2\t)+p-4)
\labell{dilsource}
\eeq
where we used that $\Tr(D)=2\cos(2\t)+2p-10$.
Note that the source is a constant independent of $k$ in agreement with the
expectation that the position space source in eq.~(\ref{souraction})
is a delta-function in the transverse directions.

The graviton source $S_h$ can be determined from either $h$-$h$ or $B$-$B$
scattering from the Dirichlet brane bound state. In the first, massless
$t$-channel exchange is mediated by only a graviton, while in the
second, both a graviton and dilaton propagate in the $t$-channel.
%\begin{figure}
%\centerline{\epsfxsize 4.7 truein \epsfbox {pic.eps}}
%\caption{Feynman diagrams for scattering of two (NS-NS or R-R)
%antisymmetric tensor states from a D-brane.}\labels{BB}
%\end{figure}
The corresponding amplitude for $B$-$B$ scattering is
\beq
A'_{BB}=
i\tS_h^{\mu\nu}(k)\,(\tG_h)_{\mu\nu,\lambda\rho}(k^2)\,
(\tV_{hBB})^{\lambda\rho}
+i\tS_\phi(k)\,\tG_{\phi}(k^2)\,\tV_{\phi BB}
\labell{BBamp}
\eeq
where the graviton propagator (in Feynman-like gauge ---
see \eg \cite{velt}) and the
three-point interactions are given by
\beqa
(\tG_h)_{\mu\nu,\lambda\rho}&=&-
\frac{i}{2}\left(\eta_{\mu\lambda}\eta_{\nu\rho}+
\eta_{\mu\rho}\eta_{\nu\lambda}-\frac{1}{4}\eta_{\mu\nu}
\eta_{\lambda\rho}\right)\frac{1}{k^2}
\nonumber\\
(\tV_{hBB})^{\lambda\rho}&=&-i\,2\kappa\,\left(
\frac{1}{2}\left(p_1\inn
p_2\,\eta^{\l\rho}-p_1^{\l}\,p_2^{\rho}-p_1^{\rho}\,p_2^{\l}
\right)\Tr({\pol}_1\inn{\pol}_2)\right.
\nonumber\\
&&\qquad\ -p_1\inn{\pol}_2\inn{\pol}_1\inn p_2\, \eta^{\l\rho}+
2\,p_1^{(\l}\,{\pol_2}^{\rho)}\inn\pol_1\inn p_2
+2\,{p_2}^{(\l}\,{\pol_1}^{\rho)}\inn\pol_2\inn p_1
\nonumber\\
&&\qquad\left.\ +2p_1\inn{\pol_2}^{(\l}\,{\pol_1}^{\rho)}\inn p_2
-p_1\inn p_2\,(\pol_1^{\l}\inn\pol_2^{\rho}+\pol_2^{\l}\inn\pol_1^\rho)
\vphantom{\frac{1}{2}}\right)
\nonumber\\
\tV_{\phi BB}&=&-i{\sqrt{2}\ka}\left(2p_1\inn\pol_2\inn\pol_1\inn p_2
-p_1\inn p_2\ \Tr(\pol_1\inn\pol_2)\right)
\labell{hbb}\\
\nonumber
\eeqa
where our notation is such that
 $\Tr({\pol}_1\cdot{\pol}_2)
=\pol_1{}^{\mu\nu}\,\pol_{2\nu\mu}$, $p_1\cdot\pol_2^\l=p_{1\delta}
\,\pol_2^{\delta\l}$ and ${\pol_1}^\rho\cdot p_2=
\pol_1^{\rho\delta}\,p_{2\delta}$. In extracting these three-point 
interactions from bulk action (\ref{nsaction}), we used  on-shell
properties of the Kalb-Ramond antisymmetric fields.
Now again we must compare
this result with the massless $t$-channel pole in the
string amplitude (\ref{finone}) with an appropriate choice
of polarization tensors.
Unraveling $\tS^{\mu\nu}$ from eq.~(\ref{BBamp}) is simplified
by noting that the only symmetric two-tensor available is
\beq
\tS_h^{\mu\nu}(k^2)=a(k^2)\,V_0^{\mu\nu}
+b(k^2)\,N^{\mu\nu}+c(k^2)V^{\mu\nu}+d(k^2)k^{\mu} k^{\nu}
\labell{tensor}
\eeq
given the symmetries of the scattering process.
Here, $V=V_0R^S$ and $R^S$ is symmetric part of the rotation matrix $R(2\t)$.
Replacing (\ref{dprop}), (\ref{dilsource}), (\ref{hbb}) and (\ref{tensor}) into
eq.~(\ref{BBamp})
\beqa
A'_{BB}&=&-\frac{i\ka}{k^2}\Big[\frac{}{}
\left(a+b+c-\frac{1}{2}\{a-b-c\}\Tr(D_0)-c\Tr(D^S)\right.\nonumber\\
&&\left.-\frac{1}{2}T_p\sqrt{1+\cF^2}\{2+
\Tr(D^S)\}\right)p_1\inn \pol_2\inn\pol_1\inn p_2
\nonumber\\
&&+4c\left(\frac{}{}p_1\inn D^S\inn\pol_2\inn\pol_1\inn p_2+p_1\inn\pol_2\inn
D^S\inn\pol_1\inn p_2+
p_2\inn D^S\inn\pol_1\inn\pol_2\inn p_1\right) \nonumber\\
&&+2(a-b-c)\left(\frac{}{}p_1\inn
D_0\inn\pol_2\inn\pol_1\inn p_2+p_2\inn D_0\inn\pol_1\inn\pol_2\inn p_1
+p_1\inn\pol_2\inn D_0 \inn\pol_1\inn p_2\right)
\nonumber\\
&&\left(\frac{}{}-(a-b-c)p_1\inn D_0\inn p_2+2cp_1\inn D^S\inn
p_1\right)\Tr(\pol_1\pol_2)+8d\,p_1\inn\pol_2\inn
p_1\,p_2\inn\pol_1\inn p_2 \Big]
\nonumber\\
&&-i\frac{\kappa}{4}\Big[\left(-(a+b-3c)+
\frac{1}{2}(a-b-c)\Tr(D_0)+c\Tr(D^S)+2d\,k^2\right)\Tr(\pol_1\inn\pol_2)
\nonumber\\
&&-4(a-b-c)\Tr(D_0\inn\pol_1\inn\pol_2)-
8c\Tr(D^S\inn\pol_1\inn\pol_2)-8d\,p_2\inn\pol_1\inn\pol_2\inn p_1 \nonumber\\
&&+\frac{T_p\sqrt{1+\cF^2}}{2}\{2+\Tr(D^s)\}\Tr(\pol_1\inn\pol_2) \Big]\,\,
\labell{ampBBone}
\eeqa
where $D^S$ is symmetric part of the $D$ matrix. 
The first bracket above is residue of the simple pole of D-brane amplitude
and the second bracket
contains some contact
terms which are related to quadratic terms of  source 
action (\ref{souraction}).

Inserting
the antisymmetric polarization tensors in (\ref{fintwo}), one may write
it as
\beqa
a_1^{BB}&=&-2\left(\frac{1}{2}p_1\inn D^S\inn p_1\,\Tr(\pol_1\pol_2)+p_1\inn
\pol_2\inn\pol_1\inn p_2+p_1\inn
D^S\inn\pol_2\inn\pol_1\inn p_2\right.
\nonumber\\
&&\,\,\, \left.+p_2\inn D^S\inn\pol_1\inn\pol_2\inn
p_1
+p_1\inn\pol_2\inn D^S \inn\pol_1\inn p_2\vphantom{\frac{1}{2}} \right)\,\,.
\labell{KK}\nonumber
\eeqa
Now comparing (\ref{ampBBone})
with the massless $t$-channel pole (\ref{pole})
fixes $a$, $b$,  $c$ and $d$ to be constants with
$a=c=-\frac{1}{2}T_p\sqrt{1+\cF^2}$ and  $b=d=0$ leaving
\beqa
\tS_h^{\mu\nu}&=&-\frac{1}{2}T_p\sqrt{1+\cF^2}\,(V_0^{\mu\nu}+V^{\mu\nu})\nonumber\\
&=&-T_p\sqrt{1+\cF^2}\cV^{\mu\nu}
\labell{gravsource}
\eeqa
where $\cV^{\mu\nu}={\rm diag}(-1,1/(1+\cF^2),1/(1+\cF^2),1,\ldots,1,0,\ldots,0)$ where zeros appear in the transvers directions. As a cross check,
we  use this  $\tS_h^{\mu\nu}$ to calculate  graviton-graviton scattering as
well.
%\begin{figure}
%\centerline{\epsfxsize 2.4 truein \epsfbox {ali9.eps}}
%\caption{Feynman diagram for graviton-graviton scattering from a D-brane
%}\labell{hhhone}
%\end{figure}
The corresponding amplitude for $h-h$ scattering is
\beqa
A'_{hh}&=&i\tS_h^{\mu\nu}(k)\,(\tG_h)_{\mu\nu,\lambda\rho}(k^2)\,
(\tV_{hhh})^{\lambda\rho}
\labell{hhamp}
\eeqa
where the three-point interaction is
\beqa
(\tV_{hhh})^{\lambda\rho}&=&-i\,2\kappa\,\left(
\left(\frac{}{}\frac{3}{2}p_1\inn
p_2\,\eta^{\l\rho}+p_1^{(\l}\,p_2^{\rho)}-k^{\l}\,k^{\rho}
\right)\Tr({\pol}_1\inn{\pol}_2)\right.
\nonumber\\
&&\qquad\ -p_1\inn{\pol}_2\inn{\pol}_1\inn p_2\, \eta^{\l\rho}+
2\,p_2^{(\l}\,{\pol_2}^{\rho)}\inn\pol_1\inn p_2
+2\,{p_1}^{(\l}\,{\pol_1}^{\rho)}\inn\pol_2\inn p_1
\nonumber\\
&&\qquad \ +2p_1\inn{\pol_2}^{(\l}\,{\pol_1}^{\rho)}\inn p_2
-p_1\inn p_2\,(\pol_1^{\l}\inn\pol_2^{\rho}+\pol_2^{\l}\inn\pol_1^\rho)
\nonumber\\
&&\qquad\left.\ -p_1\inn\pol_2\inn p_1\,\pol_1^{\l\rho}
-p_2\inn\pol_1\inn p_2\,\pol_2^{\l\rho}
\vphantom{\frac{}{}}\right)\ \ .
\labell{hhh}\nonumber
\eeqa
With this three-point interaction, the scattering amplitude (\ref{hhamp})
becomes
\beqa
A'_{hh}&=&\frac{i\ka}{k^2}T_p\sqrt{1+\cF^2}\Big[\frac{}{}2p_1\inn
\pol_2\inn\pol_1\inn p_2+
2p_2\inn\pol_1\inn\pol_2\inn D^S \inn p_2
\nonumber\\
&&+2p_1\inn\pol_2\inn\pol_1\inn D^S \inn p_1+2p_1\inn\pol_2\inn
D^S\inn\pol_1\inn p_2-p_1\inn D^S\inn p_1 \Tr(\pol_1\pol_2)
\nonumber\\
&&-\Tr(\pol_1\inn D^S)\,p_1\inn\pol_2\inn p_1-\Tr(\pol_2\inn
D^S)\,p_2\inn\pol_1\inn p_2
\nonumber\\
&&-k^2\Tr(\pol_1\inn\pol_2)-k^2\Tr(D^S\inn\pol_1\inn\pol_2)
\vphantom{\frac{}{}}\Big]\ \
\labell{hhampone}
\eeqa
Here again the terms in the last line above are contact terms and the other
terms are residue of
simple pole. Now the insertion of the graviton polarization tensor into $a_1$
gives
\beqa
a_1^{hh}&=&-2p_1\inn \pol_2\inn\pol_1\inn
p_2-2p_2\inn\pol_1\inn\pol_2\inn D^S \inn p_2-
2p_1\inn\pol_2\inn\pol_1\inn D^S \inn p_1
\nonumber\\
&&-2p_1\inn\pol_2\inn
D^S\inn\pol_1\inn p_2+p_1\inn D^S\inn p_1 \Tr(\pol_1\pol_2)
\nonumber\\
&&+\Tr(\pol_1\inn D^S)\,p_1\inn\pol_2\inn p_1+\Tr(\pol_2\inn
D^S)\,p_2\inn\pol_1\inn p_2 \ \ .
\labell{GG}
\nonumber
\eeqa
Comparing (\ref{hhampone}) with (\ref{pole}), one finds precise 
agreement between field
theory calculation
and low energy string scattering amplitude.
One can also calculate dilaton-dilaton scattering
in which the $t$-channel interaction is mediated by a graviton and the result
is consistent with the result in eq.~(\ref{gravsource}). Note however,
this amplitude alone would  not have enough structure to completely fix
all of the unknown functions appearing in eq.~(\ref{tensor}).

To determine the antisymmetric tensor source, one can  consider
$B$-$\phi$ or $B$-$h$ scattering. The scattering amplitude for $B$-$h$ is
\beqa
A'_{Bh}&=&i\tS_B^{\mu\nu}\,(\tG_B)_{\mu\nu,\l\rho}\,(\tV_{BBh})^{\l\rho}
\labell{aBh}
\eeqa
where
\beqa
(\tG_B)_{\mu\nu,\l\rho}&=&-\frac{i}{2}(\eta_{\mu\l}\eta_{\nu\rho}
-\eta_{\mu\rho}
\eta_{\nu\l})\frac{1}{k^2}
\nonumber\\
(\tV_{BBh})^{\l\rho}&=&-i2\kappa\left(p_2\inn\pol_1\inn
p_2\pol_2^{\l\rho}+2p_2\inn\pol_1^{[\l}
\pol_2^{\rho]}\inn p_1+2p_2\inn\pol_1\inn\pol_2^{[\l}p_2^{\rho]}\right.
\nonumber\\
&&\left.\,\,\,+2p_1\inn
p_2\pol_2^{[\l}\inn\pol_1{}^{\rho]}+p_2^{[\l}\pol_1^{\rho]}\inn\pol_2\inn p_1+
p_1\inn\pol_2\inn\pol_1^{[\l}p_2^{\rho]}\right)\,\,.
\nonumber
\eeqa
Here, $\pol_1$ and $\pol_2$ are graviton and the Kalb-Ramond polarization 
tensors, respectively. Since $(D^A)^{\mu\nu}$, antisymmetric part of 
the D-matrix, is the only
available antisymmetric tensor, one may write the
antisymmetric tensor source as
\beqa
\tS_B^{\mu\nu}&=&a(k^2)(D^A)^{\mu\nu}
\nonumber
\eeqa
Replacing this $B^{\mu\nu}$ source and above vertex and propagator in
(\ref{aBh})
one finds
\beqa
A'_{Bh}&=&\frac{ 2i\kappa
a}{k^2}\left(p_2\inn\pol_1\inn\pol_2\Tr(D^A\inn\pol_2)-
2p_2\inn\pol_1\inn D^A\inn\pol_2\inn p_1-2p_2\inn\pol_1\inn\pol_2\inn D^A\inn
p_2\right.
\nonumber\\
&&\left. +2p_1\inn\pol_2\inn\pol_1\inn D^A\inn p_2+k^2\Tr(\pol_2\inn\pol_1\inn
D^A)\right)
\labell{ABh}
\eeqa
Now $a_1$ for antisymmetric Kalb-Ramond and graviton is
\beqa
a_1^{Bh}&=&\Tr(\pol_2\inn D^A)p_2\inn\pol_1\inn p_2-2p_1\inn\pol_2\inn
D^A\inn\pol_1\inn p_2
\nonumber\\
&&+2p_1\inn\pol_2\inn\pol_1\inn D^A\inn p_1-2p_2\inn\pol_1\inn\pol_2\inn
D^A\inn p_2
\,\, .\nonumber
\eeqa
Replacing it in (\ref{pole}) and comparing with (\ref{ABh}) fixes  $a$ to be
constant
with $a=-\frac{1}{2}T_p\sqrt{1+\cF^2}$ leaving
\beqa
\tS_B^{\mu\nu}&=&-\frac{1}{2}T_p\sqrt{1+\cF^2}(D^A)^{\mu\nu}\nonumber\\
&=&T_p\frac{\cF^{\mu\nu}}{\sqrt{1+\cF^2}}\,\,.
\labell{Bsource}
\eeqa
 From here one can see that if $\cF=0$,  source of $B^{\mu\nu}$ field vanishes
and one would get the result of \cite{ours}.

\subsection{R-R source} \labels{rrr}

We would also like to determine source of the  Ramond-Ramond potential fields.
The relevant terms in the
low energy effective action are
\beq
I^{R-R}=\int d^{10}\!x\,\sqrt{-g}\,\sum_n\,\left(
-\frac{8}{n!}
\,F_{(n)}\inn F_{(n)}\,e^{(5-n)\frac{\ka}{\sqrt{2}}\phi}\right) \ \ .
\labell{raction}
\eeq
For the type IIb superstring, the sum runs over $n=1,$ 3 and 5,
while for the type IIa theory, the sum includes $n=2$ and 4.
An added complication is that $F_{(5)}$ should be a self-dual
field strength for which no covariant action exists.
The above non-self-dual action will yield the correct type IIb equations
of motion when the self-duality constraint is imposed by hand \cite{boon}
--- \ie one makes the substitution $F_{(5)} \rightarrow F_{(5)}+
*\!F_{(5)}$ in the equations of motion. One can verify that this action
(\ref{raction}) reproduces
the three-point string amplitudes on the sphere for two R-R vertex 
operators (\ref{vertspinor})
scattering with a graviton or dilaton (\ref{vertone}).

Now we supplement the above low energy action with the following source
term
\beqa
I^{R-R}_{\rm source}&=&\int d^{10}x[S_C^{\mu_1\cdots\mu_n}C_{\mu_1\cdots
\mu_n}]
\labell{Csource}
\eeqa
where $S_C$ is source of the R-R form potential $C_{(n)}$. 
This potential field
is defined as $dC_{(n)}=F_{(n+1)}$ for ``electric'' 
 and $dC_{(8-n)}
=*\!F_{(n+1)}$  for ``magnetic'' components of R-R 
field strength $F_{(n+1)}$. 
Here $*\!F_{(n+1)}$ is also defined by
\beqa
(F_{(n+1)})_{\mu_1\cdots\mu_{n+1}}&=&
(*\!*\!F_{(n+1)})_{\mu_1\cdots\mu_{n+1}}\nonumber\\
&=&\frac{1}{(9-n)!}(*\!F_{(n+1)})^{\nu_1\cdots\nu_{9-n}}
\epsilon^{10}_{\nu_1\cdots\nu_{9-n}\mu_1\cdots\mu_{n+1}}
\nonumber
\eeqa
To evaluate the R-R source $S_C$, we consider a scattering process
in which an external R-R field is converted to graviton. The field theory 
amplitude may be written as
\beqa
A'_{F_{(n+1)}h}&=&i(\tS_C(k))^{\mu_1\cdots\mu_n}(\tG_C(k^2))_{\mu_1\cdots\mu_n}
{}^{\nu_1\cdots\nu_n}(\tV_{CF_{(n+1)}h})_{\nu_1\cdots\nu_n}\,\,.
\labell{Afh}
\eeqa
Here  we wrote the external R-R potential field as its field strength. 
This will be convenient latter on for comparing with string amplitude. 
Now using action (\ref{raction}),  the R-R propagator and the 
three-point interactions involving ``electric'' and ``magnetic''
fields are found to be
\beqa
(\tG_C(k^2))_{\mu_1\cdots\mu_n}{}^{\nu_1\cdots\nu_n}&=&-\frac{in!}{16k^2}
\eta^{[\nu_1}_{[\mu_1}\eta^{\nu_2}_{\mu_2}\cdots\eta^{\nu_n]}_{\mu_n]}
\nonumber\\
(\tV_{C_{(n)}F_{(n+1)}h})_{\nu_1\cdots\nu_n}&=&-\frac{32\ka}{n!}
[\pol_2^{\l}{}_{\mu}
(F_{1(n+1)})_{\l\nu_1\cdots\nu_n}k^{\mu}-n\pol_2^{\l}{}_{\nu_1}(F_{1(n+1)})_{
\l\nu_2\cdots\nu_n\mu}k^{\mu}]
\nonumber\\
(\tV_{C_{(8-n)}(*\!F_{(n+1)})h})_{\nu_1\cdots\nu_{8-n}}&=&
-\frac{32\ka}{(8-n)!}[\pol_2^{\l}{}_{\mu}
(*\!F_{1(n+1)})_{\l\nu_1\cdots\nu_{8-n}}k^{\mu}\nonumber\\
&&\,\,\,\,\,\,\,\,\,-(8-n)\pol_2^{\l}{}_{\nu_1}
(*\!F_{1(n+1)})_{\l\nu_2\cdots\nu_{8-n}\mu}k^{\mu}]
\nonumber
\eeqa
where  $\pol_2$ and $F_{1(n+1)}$  are the graviton
polarization and the external R-R particle's (linearized) field strength,
 respectively. 
Now, since R-R potential is a total antisymmetric field, 
its source $S_C$ must also
be a total antisymmetric tensor. Using this and
 replacing above propagator and three-point
interactions in (\ref{Afh}), one obtains
\beqa
A'_{F_{(n+1)}h}&=&-\frac{2\ka(n+1)}{k^2}\pol_2^{\l\mu}
(F_{1(n+1)})_{\l}{}^{\nu_1\cdots\nu_n}
k_{[\nu_n}(\tS_C)_{\mu\nu_1\cdots\nu_{n-1}]}
\labell{AFhelectric}\nonumber
\eeqa
for external ``electric'' field and
\beqa
A'_{*\!F_{(n+1)}h}&=&-\frac{2\ka(9-n)}{k^2}\pol_2^{\l\mu}
(*\!F_{1(n+1)})_{\l}{}^
{\nu_1\cdots\nu_{8-n}}k_{[\nu_{8-n}}(\tS_C)_{\mu\nu_1\cdots\nu_{7-n}]}
\labell{AFhmagnetic}\nonumber
\eeqa
for external ``magnetic'' R-R field strength.
The massless $t$-channel pole for the string scattering amplitude
$A_{hF}$ comes from $a^{R-R,NS-NS}_1$ in eq.~(\ref{kinrrnn}) which is
\beqa
a_1^{R-R,h}&=&-\frac{i}{2\sqrt{2}}\Tr(P_-\G_{1(n)}M^p\g^{\nu}\g\inn
k\g^{\mu})(\pol_2\inn D)_{\mu\nu}
\nonumber\\
&=&\frac{i}{2\sqrt{2}}\Tr(\G_{1(n)}\g^{\mu}\g\inn k M^pP_+\g^{\nu})
\pol_{2\mu\nu}
\labell{aRRG}
\eeqa
where in the second line above eqs.~(\ref{Mmatrix}) and (\ref{momentum})
have been used. Using
(\ref{self}) and (\ref{finalm}), one may write eq.~(\ref{aRRG}) as
\beqa
a_1^{R-R,h}&=&
\frac{i}{4\sqrt{2}}\frac{a_nb_p}{n!(p+1)!}\,
\pol_{2\mu\nu}
F_{1\mu_1\mu_2\cdots\mu_n}\,(\e^v)_{\nu_0\nu_1\cdots\nu_{p'}}\,k_{\a} 
\nonumber\\
&&\times\{
Tr[\g^{\mu_1}\g^{\mu_2}\cdots\g^{\mu_n}\g^{\mu}\g^{\a}\g^{\nu_0}
\g^{\nu_1}\cdots
\g^{\nu_{p'}}(1+\g_{11})\g^{\nu}]
\nonumber\\
&&\qquad\,\qquad\times (\delta_{p',p}\cos\t-\delta_{p',p-2}\sin\t)\}
\labell{aRRGone}
\eeqa
where $b_p=\pm i$ for $p+1=$ odd and $b_p=\pm 1$ for $p+1=$ even. Now one
can perform the traces above for different $n$ and $p$. The result is
\[ 
a^{R-R,h}_1=\pm
i\frac{8\sqrt{2}}{n!}\pol_{2}^{\l\mu}(F_{1(n+1)}
)_\l{}^{\nu_1\cdots\nu_n}\times\left\lbrace\matrix{
(n+1)k_{[\nu_n}(\e^v)_{\mu\nu_1\cdots\nu_{n-1}]}
(\delta_{p,n-1}\cos\t+\delta_{p,n+1}\sin\t)
&\cr
k^{\rho}\,(\e^n)_{\rho\mu\nu_1\cdots\nu_n}
(\delta_{p,7-n}\cos\t+\delta_{p,9-n}\sin\t)
&\cr}
\right.
\]
where the $\pm$ sign is the same as that appearing in (\ref{finalm}).
Here $\e^v$($\e^n$) is  volume form in
subspace  parallel(orthogonal) to the brane's world-volume. Using identity
\beqa
(\e^n)_{\mu_{p+1}\cdots\mu_9}&=&-\frac{1}{(p+1)!}(\e^v)^{\mu_0\cdots\mu_p}
\e^{10}_{\mu_0\cdots\mu_p\mu_{p+1}\cdots\mu_9}
\nonumber
\eeqa
one may write $a^{R-R,h}_1$ as
\beqa
a^{R-R,h}_1&=&\pm\,i\frac{8\sqrt{2}(n+1)}{n!}\,\pol_2^{\l\mu}\,(F_{1(n+1)})_{\l}
{}^{\nu_1\cdots\nu_n}\,k_{[\nu_n}(\e^v)_{\mu\nu_1\cdots\nu_{n-1}]}
(\delta_{p,n-1}\cos\t+\delta_{p,n+1}\sin\t)
\nonumber\\
&&\pm\,i\frac{8\sqrt{2}(9-n)}{(8-n)!}\,\pol_2^{\l\mu}\,(*F_{1(n+1)})_{\l}
{}^{\nu_1\cdots\nu_{8-n}}\,k_{[\nu_{8-n}}(\e^v)_{\mu\nu_1\cdots\nu_{7-n}]}
\nonumber\\
&&\qquad\qquad\,\,\,\,\,\,\,\, \times
(\delta_{p,7-n}\cos\t+\delta_{p,9-n}\sin\t)
\labell{arrg}\nonumber
\eeqa
Now comparing (\ref{pole}) with $A'_{F_{(n+1)}h}$, one finds the R-R ``electric''
sources as
\beqa
(\tS_C)^E_{\mu_1\cdots\mu_n}&=&-\frac{4}{n!}Q_p\sqrt{1+\cF^2}(\delta_{p,n-1}
\cos\t+\delta_{p,n+1}\sin\t)(\e^v)_{\mu_1\cdots\mu_n}
\labell{sce}
\eeqa
and the ``magnetic'' sources as
\beqa
(\tS_C)^M_{\mu_1\cdots\mu_{8-n}}&=&-\frac{4}{(8-n)!}Q_p\sqrt{1+\cF^2}
(\delta_{p,7-n}
\cos\t+\delta_{p,9-n}\sin\t)(\e^v)_{\mu_1\cdots\mu_{8-n}}\,\,.
\labell{scm}
\eeqa
where $Q_p=\pm\sqrt{2}T_p$. 
In the case $\cF=0$ ($\t=0$), eq.~(\ref{sce}) indicates that brans with
$p=-1,\,0,\,1,\,2$ and 3 carry  ``electric'' charge of $C_{(n)}$ with
 $n=0,\,1,\,2,\,3$
and 4, respectively. While eq.~(\ref{scm}) shows that D$p$-branes with
$p=3,\,4,\,5,\,6$ and 7 carry ``magnetic'' charge of $C_{(n)}$ with 
 $n=4,\,5,\,6,\,7$ 
and 8, respectively. As expected the D3-brane simultaneously carries ``electric''
and ``magnetic'' charge of the self-dual $F_{(5)}$ form.  
  
\subsection{Long range fields}\labels{long}

{}From the dilaton,  graviton,  antisymmetric Kalb-Ramond and Ramond-Ramond
  sources, it is a
simple matter to
calculate (the Fourier transform of) the corresponding long range
fields around the D$p$-branes. These fields are precisely the product
of the source and the Feynman propagator in the transverse momentum
space. Hence the long range dilaton field is
\beqa
\tilde{\phi}(k^2)&=&i\tS_\phi\,\tG_{\phi}(k^2)=
-\frac{T_p\sqrt{1+\cF^2}}{4\sqrt{2}}\,\frac{2+\Tr(D)}{k^2}
\nonumber\\
&=&-\frac{T_p\sqrt{1+\cF^2}}{2\sqrt{2}}\,\frac{\cos(2\t)+p-4}{k^2}\ \ .
\labell{dilfield}
\eeqa
Similarly the long range gravitational field becomes
\beqa
\tilde{h}_{\mu\nu}(k^2)&=&i\tS_h^{\l\rho}\,(\tG_h)_{\l\rho,\mu\nu}
\nonumber\\
&=&-\frac{T_p\sqrt{1+\cF^2}}{8k^2}{\rm
diag}\left(-\a_p,\g_p,\g_p,\a_p,\cdots,\a_p,\b_p,\cdots,\b_p\right)
\labell{gravfield}
\eeqa
where $\a_p=-p+9-2\cos^2\t$, $\g_p=6\cos^2\t-p+1$ and $\b_p=-p+1-2\cos^2\t$.
The long range antisymmetric
tensor field is
\beqa
\tB_{\mu\nu}(k^2)&=&i\tS_B^{\l\rho}(\tG_B)_{\l\rho,\mu\nu}
\nonumber\\
&=&-\frac{T_p}{2k^2}\sqrt{1+\cF^2}(D^A)_{\mu\nu}\nonumber\\
&=&\frac{T_p}{2k^2}\sqrt{1+\cF^2}\sin(2\t)(\epsilon^2)_{\mu\nu}
\labell{Bfield}
\eeqa
where $\epsilon^2$ is volume form in the $(x^1,x^2)$-plane.
%\begin{figure}
%\centerline{\epsfxsize 2.4 truein \epsfbox {p4.eps}}
%\caption{Feynman diagram for graviton--R-R tensor scattering from a D-brane}
%\end{figure}
Finally, the long range R-R potential fields are
\beqa
\tC_{\mu_1\cdots\mu_n}(k^2)&=&i(\tS_C)^{\nu_1\cdots\nu_n}
(\tG_C)_{\nu_1\cdots\nu_n,
\mu_1\cdots\mu_n}\nonumber\\
&=&\frac{n!}{16k^2}(\tS_C)_{\mu_1\cdots\mu_n}\nonumber\\
&=&-\frac{Q_p}{4k^2}\sqrt{1+\cF^2}[\cos\t\delta_{p,n-1}+\sin\t\delta_{p,n+1}]
(\e^v)_{\mu_1\cdots\mu_n}
\labell{rrfield}
\eeqa
From here one can see that  a given $C_{(n)}$ simultaneously couples to 
both D$(p+1)$- and D$(p-1)$-branes.

Now we would like to compare above long range fields with the low-energy
background field solutions corresponding to $(p+1)$- and $(p-1)$-brane 
bound states\cite{breck}. For simplicity we consider only the 
2-brane and 0-brane 
bound state solution. In \cite{breck} this solution was found, by rotating a
delocalized 1-brane in $(x^1,x^2)$-plane and using T-duality on 
the rotated brane,
to be (in the Einsten-frame)
\beqa
d\hs^2&=&e^{-\hphi/2}\sqrt{H}\Big\{\frac{-dt^2}{H}+\frac{(dx^1)^2
+(dx^2)^2}{1+(H-1)\cos^2\t}
+\sum_{i=3}^9(dx^i)^2\Big\}\nonumber\\
\hC_{(3)}&=&\pm\frac{(H-1)\cos\t}{1+(H-1)\cos^2\t}
dt\wedge dx^1\wedge dx^2
\nonumber\\
\hC_{(1)}&=&\pm\frac{H-1}{H}\sin\t dt\nonumber\\
\hB&=&\frac{(H-1)\cos\t\sin\t}{1+(H-1)\cos^2\t}dx^1\wedge dx^2\nonumber\\
e^{2\hphi}&=&\frac{H^{\frac{3}{2}}}{1+(H-1)\cos^2\t}
\labell{lowsolution}
\eeqa
where $H=1+\mu G(r/\cl)$ and $G(r/\cl)=\frac{1}{5}(\cl/r)^5$ is the 
Green's function
in the transverse space.
Here $r^2=\sum_{i=3}^9(x^i)^2$ and $\cl$ is an arbitrary length. 
Also $\mu$ is some dimensionless constant and we consider it
 to be small, \ie $\mu<<1$, so that the 
nontrivial part of above solution may be treated as a perturbation of flat 
empty space. The perturbative parts, \ie $\mu<<1$, or long range fields,
\ie $r\longrightarrow \infty$,  of above solution  are
\beqa
\hh_{\mu\nu}&\equiv& \hat{g}_{\mu\nu}-\eta_{\mu\nu}\nonumber\\
&\simeq&-\frac{\mu G}{8}{\rm diag}(-\a_2,\g_2,\g_2,\b_2,\b_2,
\cdots,\b_2)  \nonumber\\
(\hC_{(3)})_{\mu_1\mu_2\mu_3}&\simeq&\pm\mu G\cos\t(\e^v)_
{\mu_1\mu_2\mu_3}\nonumber\\
(\hC_{(1)})_{\mu}&\simeq&\pm\mu G\sin\t(\e^v)_{\mu}
\nonumber\\
\hB_{\mu\nu}&\simeq&\mu G\cos\t\sin\t(\epsilon^2)_{\mu\nu}
\nonumber\\
\hphi&\simeq&-\frac{\mu G}{4}(2\cos^2\t-3)
\labell{perlow}
\eeqa
where $\a_2=7-2\cos^2\t$, $\g_2=6\cos^2\t-1$ and $\b_2=-1-2\cos^2\t$.  To make
a precise comparison of the long range fields (\ref{dilfield}--\ref{rrfield}), 
with above asymptotic fields, we must first
take into account that the low energy string actions, (\ref{nsaction}) and
(\ref{raction}), have different normalization from that used in \cite{breck}
to derive the 
bound state solution (\ref{lowsolution}). The latter action  would have
the same normalization as the former  if one
uses the following field redefinitions
\beqa
\hh_{\mu\nu}&\equiv&2\ka h'_{\mu\nu}\nonumber\\
\hphi&\equiv&\sqrt{2}\ka\phi'\nonumber\\
\hB_{\mu\nu}&\equiv&- 2\ka B'_{\mu\nu}\nonumber\\
\hC_{(n)}&\equiv&4\sqrt{2}\ka C'_{(n)}
\labell{prime}
\eeqa
We have to also make Fourier transform of the asymptotic fields (\ref{perlow})
in order
to compare them with the momentum space  long range fields (\ref{dilfield}--
\ref{rrfield}). The essential transform is that for $G(r/\cl)$ which is
\[
\tG(k^2)\,=\,\frac{\cA_6\cl^5}{k^2}
\]
where ${\vec k}$ is a wave-vector in the subspace orthogonal to the 2-brane,
and $\cA_6$ is the area of a unit 6-sphere. Now in terms of the primed fields
introduced in eq.~(\ref{prime}), the Fourier transform of the asymptotic fields
(\ref{perlow}) are  
\beqa
\tilde{h}'_{\mu\nu}&\simeq&-\frac{\mu\cA_6\cl^5}{16\ka}\frac{1}{k^2}{\rm diag}
(-\a_2,\g_2,\g_2,\b_2\cdots,\b_2) \nonumber\\
\tphi'&\simeq&-\frac{\mu\cA_6\cl^5}{4\sqrt{2}\ka}\frac{1}{k^2}(2\cos^2\t-3)
\nonumber\\
\tilde{B}'_{\mu\nu}&\simeq&-\frac{\mu\cA_6\cl^5}{2\ka}\frac{1}{k^2}\cos\t\sin\t
(\e^2)_{\mu\nu}\nonumber\\
(\tilde{C}'_{(3)})_{\mu_1\mu_2\mu_3}&\simeq&\pm\frac{\mu\cA_6\cl^5}
{4\sqrt{2}\ka}
\frac{1}{k^2}\cos\t(\e^v)_{\mu_1\mu_2\mu_3}\nonumber\\
(\tilde{C}'_{(1)})_{\mu}&\simeq&\pm\frac{\mu\cA_6\cl^5}{4\sqrt{2}\ka}
\frac{1}{k^2}
\sin\t(\e^v)_{\mu}\,\,. \nonumber
\eeqa
Hence we have complete agreement between the D-brane long range fields 
(\ref{dilfield}--\ref{rrfield}) and those above if $2\ka T_p\sqrt{1+\cF^2}
=\mu\cA_6\cl^5$. 
In \cite{breck}, by evaluating the long range fields, mass and R-R charge of
the $(p+1)$- and $(p-1)$-brane bound state was found and shown that they satisfy
extremal condition.
Therefore, the long range fields (\ref{dilfield}--\ref{rrfield})  are fields
around extremally charged D$(p+1)$- and D$(p-1)$-brane bound state indicating as expected that 
the mixed
boundary condition is world-sheet realization of the D-brane bound states.

\subsection{D-brane action}\labels{dbraneaction}

In this section we show that the source actions derived from scattering
amplitude
of  sect.~(\ref{rest})
are  consistent with Born-Infeld and
Chern-Simons actions. To this end consider first the Born-Infeld action
\footnote{We are not interested in this paper in 
coupling of open string scalar and gauge fields to the 
D-brane.}
\beqa
S_{BI}&=&-T_p\int d^{p+1}x\,e^{-\Phi}\sqrt{-{\rm det}(G_{ab}+\tilde{\cF}_{ab})}
\nonumber
\eeqa
where $G_{ab}$ and $\Phi$ are  the string-frame metric and dilaton. Here
$\tilde{\cF}_{ab}$ contains
background field and its  quantum fluctuation. In order to compare above action
with our result, one should
write the Born-Infeld action in terms of the Einstein-frame
 metric and dilaton with the normalizations of the low energy 
action (\ref{nsaction}). Hence replacing
$G_{ab}=e^{\Phi/2}g_{ab}$ and $\Phi=\sqrt{2}\phi$, one finds
\beqa
S_{BI}&=&-T_p\int d^{p+1}x\,e^{\frac{p-3}{2\sqrt{2}}\phi}\sqrt{-{\rm
det}(g_{ab}+e^{-\frac{\phi}{\sqrt{2}}}\tilde{\cF}_{ab})}
\nonumber
\eeqa
Now we expand this action around backgrounds $\eta_{ab}$ and $\cF_{ab}$ 
which appear
in $g_{ab}=\eta_{ab}+2h_{ab}$ and $\tilde{\cF}_{ab}=\cF_{ab}-2B_{ab}$, and keep
only terms that are linear in $h_{ab}$, $\phi$ and $B_{ab}$. Using  the
following expansion
\beqa
\sqrt{{\rm det}(A+\delta A)}&=&\sqrt{{\rm
det}(A)}[1+\frac{1}{2}\Tr(A^{-1}\delta A)+\cdots ] \nonumber
\eeqa
and  identities
\beqa
\sqrt{-{\rm
det}(\eta_{ab}+e^{-\frac{\phi}{\sqrt{2}}}\cF^{ab})}&=&\sqrt{1+
e^{-\frac{2\phi}{\sqrt{2}}}\cF^2}\nonumber\\
 (\eta_{ab}+\cF_{ab})^{-1}&=&\cV_{ab}-\frac{\cF_{ab}}{1+\cF^2}
\nonumber
\eeqa
one finds
\beqa
S_{BI}&=&-T_p\sqrt{1+\cF^2}\int d^{p+1}x \Big[1+\cV^{ab}h_{ab}+
(\frac{p-3}{2}-
\frac{\cF^2}{1+\cF^2})\frac{\phi}{\sqrt{2}}\nonumber\\
&&\qquad\qquad+\frac{\cF^{ab}B_{ba}}{1+\cF^2}+\cdots \Big]
\labell{biaction}
\eeqa
where the dots represent the linear coupling of
open string fields to D-brane which can not be evaluated in our calculations,
as well as higher order couplings. Here the indices are raised and lowered by
$\eta^{ab}$ and $\eta_{ab}$, respectively. 

The Fourier transform sources  (\ref{dilsource}), (\ref{gravsource}),
(\ref{Bsource}),  (\ref{sce}) and (\ref{scm})
are constant, so they appear as $S=\tS\delta(x^i)$ in position space.
Now replacing the NS sources   in (\ref{souraction}), one finds
\beqa
I^{NS-NS}_{source}&=&
 -T_p\sqrt{1+\cF^2}\int
d^{p+1}x[\cV^{ab}h_{ab}+\frac{1}{\sqrt{2}}(\frac{p-3}{2}
-\frac{\cF^2}{1+\cF^2})\phi
-\frac{\cF^{ab}B_{ab}}{1+\cF^2}]
\nonumber
\eeqa
As can be seen, this action is the linear part of the Born-Infeld 
action (\ref{biaction})

The Chern-Simons action is
\beqa
S_{CS}&=&-\mu_p\int e^{\tilde{\cF}}\sum c_{(p+1)}\nonumber\\
&=&-\mu_p\int\,dx^{p+1}[\frac{1}{(p+1)!}c^{\mu_0\cdots\mu_p}
+\frac{1}{(p-1)!2!}
c^{\mu_0\cdots\mu_{p-2}}\tilde{\cF}^{\mu_{p-1}\mu_p} \nonumber\\
&&+\frac{1}{(p-3)!2!2!2!}c^{\mu_0\cdots\mu_{p-4}}
\tilde{\cF}^{\mu_{p-3}\mu_{p-2}}
\tilde{\cF}^{\mu_{p-1}\mu_p}+\cdots](\e^v)_{\mu_0\cdots\mu_p}
\nonumber
\eeqa
where $c_{(n)}$ is the R-R form potential that has different normalization
from that used in bulk action (\ref{raction}). Our calculations are
consistent if one sets $c_{(n)}=4C_{(n)}$. 
For special case of $\tilde{\cF}_{\mu\nu}=\cF_{\mu\nu}$ which has only two
components, \ie $\cF_{12}$ and $\cF_{21}$, above action simplifies to
\beqa
S_{CS}
&=&-4\mu_p\int\,d^{p+1}x[\frac{1}{(p+1)!}C^{\mu_0\cdots\mu_p}+
\frac{1}{(p-1)!2!}
C^{\mu_0\cdots\mu_{p-2}}\cF^{\mu_{p-1}\mu_p}
](\e^v)_{\mu_0\cdots\mu_p}
\labell{sCS}
\eeqa
Now writing sources (\ref{sce}) and (\ref{scm}) in position 
space and replacing
them in (\ref{Csource}), one finds
\beqa
I^{R-R}_{source}&=&-4Q_p\int d^{p+1}x\sqrt{1+\cF^2}\nonumber\\
&&\,\,\,\,\,\,\,\,\times[\frac{1}{(p+1)!}C^{\mu_0\cdots
\mu_p}\cos\t+\frac{1}{(p-1)!2!}C^{\mu_0\cdots
\mu_{p-2}}(\epsilon^2)^{\mu_{p-1}\mu_p}\sin\t](\epsilon^v)_{\mu_0\cdots\mu_p}
\nonumber\\
&=&-4Q_p\int d^{p+1}x[\frac{1}{(p+1)!}C^{\mu_0\cdots
\mu_p}+\frac{1}{(p-1)!2!}C^{\mu_0\cdots
\mu_{p-2}}\cF^{\mu_{p-1}\mu_p}](\epsilon^v)_{\mu_0\cdots \mu_p}
\labell{irrone}\nonumber
\eeqa
where  we used (\ref{cossin}) to replace $\cos\t$ and
$\sin\t$ in terms of $\cF$.
This action is actually the Chern-Simons action (\ref{sCS})  
 by setting $\mu_p=Q_p$.

\section{Discussion} \labels{discuss}

In this paper, we have presented detailed calculations of all
two-point amplitudes describing massless closed type II superstrings
scattering from a
Dirichlet $p$-branes that carry a background magnetic field. 
Using these results we
derived the long
range fields around the branes  
 and showed as expected  they
are
field around extremally charged D$(p)$- and D$(p-2)$-brane bound states for
$2\le p\le 7$. From the scattering
amplitude
we also calculated linear coupling of closed string fields
to the D-brane bound states and found that they are
consistent
with Born-Infeld and Chern-Simons actions.

The low-energy background field solutions corresponding  to D-brane
bound states that presented in \cite{breck} possess one half of the 
supersymmetries of type II superstring theories. This happened because
those solutions were constructed by imposing T-duality map on a single
D-brane solution of type II theories. Supersymmetry is preserved by
T-duality, hence the bound state solutions retain the supersymmetric
properties of the initial (single) D-brane.
In conformal field theory frame, as discussed in \cite{ours}, any scattering procces that is describable  in terms of $D$ and $M$ matrices has the same supersymmetry properties as type I theory.  
So  (single) D-branes as well as the   D-brane bound states that studied in this paper posses one half of the type II
supersymmetries.

Using the ADM mass formula, mass per unit $p$-volume of the 
D$p$- and D$(p-2)$-brane 
bound state is $M_{p,p-2}=T_p\sqrt{1+\cF^2}/\ka\equiv T_{p,p-2}/\ka$.
Hence the D-brane bound states has tension $T_{p,p-2}=T_p\sqrt{1+\cF^2}$,
the one that used in the scattering amplitude (\ref{ampuniverse}). 
Similarly, R-R charges   per unit $p$-volume are $Q_p=\pm\sqrt{2}T_p$
which was used in eqs.~(\ref{sce}) and (\ref{scm}), 
and $Q_{p-2}=\pm\sqrt{2}T_p\cF$. Therefore, they satisfy
the extremal condition $(Q_p^2+Q_{p-2}^2)=2T_{p,p-2}^2$ \cite{breck}.

{}From the gamma function factors appearing in eq.~(\ref{finone}), we see that
the amplitudes contain two infinite series poles corresponding to on-shell
propagation of closed and open strings in $t$- and $q^2$-channels, 
respectively.
The $t$-channel poles (\ie $\alpha' t=4n$ with $n=0,1,2,\ldots$) indicate that
the mass of closed strings are $(m^{closed})^2=4n/\alpha'$ because  
$t=-(p_1+p_2)^{\mu}(p_1+p_2)^{\nu}\eta_{\mu\nu}=(m^{closed})^2$ 
where $(p_1+p_2)^{\mu}$ 
is 
momentum of the closed strings. The $q^2$- or $s$-channel poles
(\ie $\alpha' q^2=-n$) indicate that the open string masses are
$(m^{open})^2=n/\alpha'$. This is a result of the relation 
$q^2=p_1^ap_1^b\cV_{ab}
=-(m^{open})^2$ where $p_1^a=(p_1\inn V_0)^{\mu}$ is the open string momentum.
Hence the string tension seems to be shifted in an anisotropic manner on
the D-brane world volume. The effect is to induce a non-trivial metric 
 on its world-volume, \ie one which  is not just
$\eta_{ab}$. Hence one might say that  
the D-brane world-volume is no longer flat. Of course, the closed strings propagating  in the bulk remain unaffected by this shift, and hence they only see flat spacetime.

While our analyses of the R-R sources in sect.~\ref{rrr} are restricted
to $2\leq p\leq 7$, the NS-NS sources  in sect.~\ref{nsnsns} 
are also valid in the case of domain walls
(for which $p-8$). 
In this case as a result of ``non-flat'' nature of the world-volume metric
$t\neq 4q^2$ and the kinematic factor
$a_1(1,2)$ is not proportional to $t$, in contrast to the kinematic 
properties of amplitudes of scattering from single D-brane \cite{ours}.
Also, the  scattering amplitudes
 in field theory side do have non-zero 
simple poles which are equal to simple poles
of string amplitude. Hence, for special case of $\cF=0$, 
 our results verify that
the NS-NS long range fields around D8-brane correspond to 
those of extremally charged
$(p=8)$-brane solution of the low energy effective action, 
things which we were
not able to show in \cite{ours}.

The  $t$-channel analyses of sect. \ref{massless} enabled 
us to find linear coupling of closed string
fields to the D-brane. The  quadratic  and higher order couplings appear as
contact interactions. In general, to find these contact terms, 
one should equate
contact terms of string amplitude to contact terms of field theory 
(with  bulk and source actions) in both
$t$- and $s$-channels. 
The two point amplitudes presented in sect. \ref{rest} has 
enough structure to
extract  the  contact terms corresponding to two closed string fields. 
However,  evaluation of these 
contact terms in field theory side 
in $s$-channel needs an understanding of coupling of
open string
gauge and scalar fields to D-brane \cite{ourstwo}. It would  be also 
interesting to
evaluate coupling of massless 
open string fields to the D-brane bound state and show that  
they are consistent with Born-Infeld and Chern-Simons actions \cite{aki}. 

Finally, although the coupling of closed string fields to the D-brane world-volume that we found in this paper can be extracted from abelian Born-Infeld and Chern-Simons actions, extending the boundary condition (\ref{mixboundary}) to includes non-abelian
 gauge fields helps one to find non-abelian Born-infeld and Chern-Simons actions \cite{nonabelian}, things which are not fully understood yet \cite{nonabelian1}.

\section*{Acknowledgments}
I gratefully aknowledge contributions of Robert C. Myers to an early stage of this work, and for suggesting improvements to a previous draft of this paper. I would also like to acknowledge useful conversations with H. Arfaei and J.C. Breckenridge.

\section*{Appendix}

In this appendix we would like to evaluate the $D$ and $M$ matrices for
moving D(p)-brane and D(p)-brane with its electric field turned on. To this end,
we begin with a D(p)-brane that move with constant velocity $v$ in one of its 
transvers direction, say $X^1$. Then appropriate boundary condition for the
open strings on this brane is
\beqa
\prt_y(X^0-vX^1)&=&0\nonumber\\
X^1-vX^0&=&0\nonumber\\
\prt_y X^a&=&0\nonumber\\
X^i&=&0\nonumber
\eeqa
where $a=2,3,...,p+1$ and $i=p+2,...,9$. Now following the steps in 
sect.~\ref{conformal}, one finds 
\beqa
(D_v^p)^{\mu\nu}&=&(D_0^{p+1})^{\mu}{}_{\alpha}R_v^{\alpha\nu}\nonumber
\eeqa
where
\[
{R_v^{\alpha\nu}}=-\pmatrix{\cosh 2\t_v & \sinh 2\t_v\cr
\sinh 2\t_v & \cosh 2\t_v\cr}
\]
and we defined $\cosh 2\t_v\equiv (1+v^2)/(1-v^2)$ and
$\sinh 2\t_v\equiv 2v/(1-v^2)$.  
The $M$ matrix is also found to be
\beqa
M_v^p&=&i\g^{11}(\g^1\cosh\t_v-\g^0\sinh\t_v)M_0^{p+1}
\nonumber
\eeqa
where $D_0^{p+1}(M_0^{p+1})$ is the appropriate $D(M)$
 matrix for stationary D(p+1)-brane which
includes $X_1$ as its world-volume.

To write the $D$ and $M$ matrices for D(p)-branes which carry constant electric field, we consider the electric field to be non-zero only in $X^1$ direction. Then appropriate boundary condition can be read from (\ref{mixboundary}) to be
\beqa
\prt_y X^0+iE\prt_x X^1&=&0\nonumber\\
\prt_y X^1+iE\prt_x X^0&=&0\nonumber\\
\prt_y X^b&=&0\nonumber\\
X^j&=&0\nonumber
\eeqa
where $b=2,3,...,p$ and $j=p+1,...,9$. From these boundary conditions, one finds the $D$ matrix as
\beqa
(D_E^p)^{\mu\nu}&=&(D_0^p)^{\mu}{}_{\alpha}R_E^{\alpha\nu}\nonumber
\eeqa
where
\[
R_E^{\alpha\nu}=-\pmatrix{\cosh 2\t_E & -\sinh 2\t_E\cr
\sinh 2\t_E & -\cosh 2\t_E\cr}
\]
and we defined $\cosh 2\t_E\equiv (1+E^2)/(1-E^2)$ and
$\sinh 2\t_E\equiv 2E/(1-E^2)$. The $M$ matrix can be evaluated and the result is 
\beqa
M_E^p&=&(\cosh\t_E+\g^0\g^1\sinh\t_E)M_0^p\nonumber
\eeqa
Note that in both cases the $D$ matrix satisfy 
$D^{\mu}{}_{\alpha}D^{\alpha\nu}=\eta^{\mu\nu}$. A difference between these two $D$ matrices is that the $(D_v^p)^{\mu\nu}$ is symmetric while the $(D_E^p)^{\mu\nu}$ is not. So in scattering amplitude of boundary states from D(p)-branes which carry electr
ic field the antisymmetric part of $(D_E^p)^{\mu\nu}$ produces some phase factors \cite{gukov}.

\end{document}